\documentclass[final,authoryear,5p,times,twocolumn]{elsarticle}

\usepackage{graphicx} 
\usepackage{amsfonts} 
\usepackage{amsmath} 
\usepackage{amssymb} 
\usepackage{hyperref} 
\usepackage{siunitx} 
\usepackage{booktabs} 
\usepackage{cleveref} 
\usepackage{color} 
\usepackage{ulem}

\providecommand{\e}[1]{\ensuremath{\times 10^{#1}}}
\providecommand{\e}[1]{\ensuremath{\times 10^{#1}}}
\crefformat{appendix}{#2#1#3}

\begin{document} 
\begin{frontmatter}
	
	\title{Quantification of systemic risk from overlapping portfolios in the financial system}
	
	\author[iiasa,hub,cosy]{Sebastian Poledna} \ead{poledna@iiasa.ac.at}
	
	\author[banco]{Seraf\'{i}n Mart\'{i}nez-Jaramillo} \ead{smartin@banxico.org.mx}
	
	\author[ucl,sr]{Fabio Caccioli} \ead{f.caccioli@ucl.ac.uk} 
	
	\author[cosy,hub,sfi,iiasa]{Stefan Thurner\corref{cor}} \ead{stefan.thurner@meduniwien.ac.at}
	
	\cortext[cor]{Corresponding author}
	
	\address[iiasa]{IIASA, Schlossplatz 1, A-2361 Laxenburg, Austria}
		
	\address[banco]{Direcci\'{o}n General de Estabilidad Financiera, Banco de M\'{e}xico, Ave. 5 de Mayo 2, Ciudad de M\'{e}xico, Distrito Federal, M\'{e}xico}
	
	\address[cosy]{Section for Science of Complex Systems, Medical University of Vienna, Spitalgasse 23, A-1090, Austria} 
		
	\address[hub]{Complexity Science Hub Vienna, Josefst{\"a}dter Stra{\ss}e 39, 1080 Vienna, Austria} 
	
	\address[sfi]{Santa Fe Institute, 1399 Hyde Park Road, Santa Fe, NM 87501, USA} 
	
	\address[ucl]{Department of Computer Science, University College London, London, WC1E 6BT, UK} 
	
	\address[sr]{Systemic Risk Centre, London School of Economics and Political Sciences, London, UK}
	
	\begin{abstract}
		Financial markets are exposed to systemic risk, the risk that a substantial fraction of the system ceases to function and collapses. Systemic risk can propagate through different mechanisms and channels of contagion. One important form of financial contagion arises from indirect interconnections between financial institutions mediated by financial markets. This indirect interconnection occurs when financial institutions invest in common assets and is referred to as {\em overlapping portfolios}. In this work we quantify systemic risk from indirect interconnections between financial institutions. Having complete information of security holdings of major Mexican financial intermediaries and the ability to uniquely identify securities in their portfolios, allows us to represent the Mexican financial system as a bipartite network of securities and financial institutions. This makes it possible to quantify systemic risk arising from overlapping portfolios. We show that focusing only on direct exposures underestimates total systemic risk levels by up to 50\%. By representing the financial system as a multi-layer network of direct exposures (default contagion) and indirect exposures (overlapping portfolios) we estimate the mutual influence of different channels of contagion. The method presented here is the first objective data-driven quantification of systemic risk on national scales that includes overlapping portfolios. 
	\end{abstract}
	
	
\end{frontmatter}

\section{Introduction} \label{intro}
Systemic risk (SR) in financial markets is the risk that a significant fraction of the financial system can no longer perform its function as a credit provider and collapses. In a more narrow sense, SR is the notion of contagion or impact that starts from the failure of a financial institution (or a group of institutions) and propagates through the financial system and potentially also to the real economy \citep{De-Bandt:2000aa,BIS:2010aa}. In a broad sense SR also includes system-wide shocks that affect many financial institutions or markets at the same time \citep{De-Bandt:2000aa}. 

SR arises through the probability of default propagating through many different mechanisms and channels of contagion. In addition to default contagion, financial SR arises from asset price shocks and funding liquidity shocks \citep{Haldane:2011aa}. Losses from asset price shocks can result in contagious failures. \emph{Marking to market}, an accounting practice of valuing assets according to current market prices, can induce a further round of assets sales, depressing prices further and lead to ``fire sales'' \citep{Cifuentes:2005aa}. Liquidity hoarding in interbank funding markets can cascade through a financial network with severe consequences \citep{Gai:2010aa,Gai:2011aa}. In \citet{Brunnermeier:2009aa} it is shown that funding of traders affects, and is affected by, market liquidity in a non-trivial way. Market liquidity and funding liquidity shocks are therefore mutually reinforcing and can lead to liquidity spirals. 

An important form of financial contagion arises from indirect links between financial institutions mediated by financial markets. When financial institutions invest in the same assets, their portfolios are said to overlap. This indirect connection is referred to as {\em overlapping portfolios}. Contagion can occur because of shocks that cause common assets to be devalued. Devaluations can cause further asset sales and devaluations leading to fire sales \citep{Cifuentes:2005aa, Huang:2013aa,Corsi:2013aa,Cont:2014aa,Caccioli:2014aa,Caccioli:2015aa}. During ``the Great Moderation'', a period starting in the mid-1980s until 2007, portfolios of financial institutions became increasingly similar. For example in 2007, before the financial crisis, many large banks around the world held mortgage-backed securities (MBS) and collateralized debt obligations (CDO) in their portfolios. During the U.S. subprime mortgage crisis these banks faced together write-downs and losses on the value of these investments. Together the losses on MBS and CDOs due to the subprime mortgage crisis totaled more than $\$500$ billion \citep{Bloomberg:2012aa}. 

Financial contagion can be studied in empirical data through the availability of high-precision financial network data. Driven by data availability, research on financial networks has mainly focused on default contagion: mostly, on direct lending networks between financial institutions \citep{Upper:2002aa,Boss:2004aa,Boss:2005aa,Soramaki:2007aa,Iori:2008aa,Cajueiro:2009aa,Bech:2010aa,Fricke:2014aa,Iori:2014aa}, but also on the network of derivative exposures \citep{Markose:2012aa,Markose:2012ab}. Research on financial \emph{multi-layer} networks, that considers multiple channels of contagion, has only appeared recently. \citet{Poledna:2015aa} and \citet{Leon:2014aa} study the interactions of financial institutions on different financial markets in Mexico and Colombia, respectively. Due to the lack of empirical data, research on networks of overlapping portfolios is practically non existing up to now. \citet{Caccioli:2014aa} study two channels of contagion by combining empirical data on direct lending with a stylized model on overlapping portfolios. \citet{Greenwood:2014aa} and \citet{Duarte:2015aa} analyze networks of common asset exposures in the EU and the US with aggregated data on asset classes.

In this context several network-based SR-measures have been proposed recently \citep{Battiston:2012aa,Markose:2012aa,Billio:2012aa,Thurner:2013aa}. These approaches bear the notion of the {\em systemic importance} of a financial institution within a financial network and rely on {\em network centrality measures} or on closely related measures. A serious disadvantage of centrality measures is that the corresponding value for a particular institution has no clear interpretation as a measure for expected losses. A breakthrough that solves this problem is the so-called DebtRank, a recursive method suggested by \citet{Battiston:2012aa} that quantifies the systemic importance of financial institutions in terms of losses that they would contribute to the total loss in a crisis. Since data on asset-liability networks is hard to obtain  outside Central Banks and is not publicly available, there have been several attempts to quantify systemic importance of institutions without the explicit knowledge of the underlying networks \citep{Cooley:2009aa,Adrian:2011aa,Acharya:2010aa,Acharya:2013aa}.

In this work we develop a novel method to quantify SR from overlapping portfolios. First, we extend the notion of systemic importance in financial networks to bipartite networks of financial institutions and securities. This makes it possible to assess SR from overlapping portfolios. Second, we compare SR from direct exposures (default contagion) and indirect exposures (overlapping portfolios). Using the methodology developed in \citet{Poledna:2015aa}, we represent the financial system as a multi-layer network and assess SR contributions from direct and indirect exposures. Third, we compare the marginal contributions of individual direct and indirect exposures to the overall SR. Building on the work of \citet{Poledna:2014aa} and  \citet{Poledna:2015aa}, we define a  novel risk measure to quantify the expected loss due to SR that takes cascading into account by explicit use of financial network topologies. We use this risk measure to quantify marginal contributions of individual direct and indirect exposures to the overall SR.

This work is based on a unique data set containing various types of daily exposures between the major Mexican financial intermediaries (banks) over the period 2004-2013 (for this work we use data from 2009-2013). The data includes detailed information of security holdings of Mexican financial intermediaries by containing the International Securities Identification Number (ISIN) that uniquely identifies every security. The data further contains the capitalization of banks at every month and the market data (pirces) for the various securities. Data were collected and are owned by the Banco de M\'{e}xico. Various aspects of the data have been studied before \citep{Martinez-Jaramillo:2010aa,Lopez-Castanon:2012aa,Martinez-Jaramillo:2014aa, Molina-Borboa:2015aa,Poledna:2015aa}. In this work we focus on the SR arising from overlapping portfolios. Having complete information of security holdings of major financial intermediaries and the ability to uniquely identify securities in the portfolios allows us to represent the Mexican financial system as a bipartite network of securities and financial institutions. With this data we quantify the SR contributions of direct exposures (default contagion) and indirect exposures (overlapping portfolios) and estimate the mutual influence of different channels of contagion.

Our paper is structured in the following way. In section 2 we explain the methodology to quantify systemic risk in bipartite and multi-layer networks. In section 3 we describe the data set used for this study. In section 4 we present the results and visualize overlapping portfolios in the Mexican financial system as a bipartite network. Finally, in section 5 we discuss the results.

\section{Quantification of systemic risk from overlapping portfolios}
\subsection{Quantification of systemic risk in bipartite networks}
We represent the financial system as a bipartite network $S_{ia}$ of banks and assets. A link connects bank $i$ to asset $a$ if $a$ is in $i$'s investment portfolio. Contagion can occur in the system whenever the same asset is shared by more than one bank. In this case a bank selling off its portfolio can cause losses to other banks with overlapping portfolios. This occurs because of market-impact, that is the tendency of prices to move in response to trading activity. In particular, the price of an asset is expected to drop by an amount that depends on the size of the position that is liquidated. It is then clear that banks with similar portfolios are mutually exposed even if there are no direct linkages between them (in form of e.g. interbank lending). A correct assessment of systemic risk should account for the stress imposed on the system by assets liquidation. 

We address this problem by considering a modification of DebtRank for bipartite networks, a methodology recently introduced to identify systemically important banks in a network of mutual exposures \citep{Battiston:2012aa}. It is a quantity that measures the fraction of the total economic value $V$ in the network that is potentially affected by the distress of an individual node (bank) $i$, or by a set of nodes $S$. The DebtRank of a set of nodes $S$ that is initially in distress is denoted by $R_{S}$. In those cases where only one node $i$ is initially under distress (the set $S$ contains only one node $i$) we denote the DebtRank of that node by $R_i$. For details see \cref{debtrank_section}. 

DebtRanks can be calculated from any financial network representing financial interdependencies \citep{Poledna:2015aa}. Although different financial interdependencies are associated with different types of financial risk, the interdependencies can be represented by an \emph{exposure network}. We use the following notation for different exposure types: the size of every exposure of type $\alpha$ of institution $i$ to institution $j$  at time $t$ is given by the matrix element $X_{ij}^{\alpha}(t)$. $\alpha=1,2$ labels the layers ``direct exposures,''  and ``indirect exposures'' respectively. DebtRank of layer $\alpha$ is given by $R_i^{\alpha}(t)=R_i^{\alpha}(t)(X_{ij}^{\alpha}(t),C_{i}(t),v^{\alpha}_{i}(t))$ with $v_{i}^{\alpha}(t)$ as the respective economic value of layer $\alpha$ at time $t$. The links between nodes have the same meaning for all financial interdependencies: it is the total loss that might arise for a bank as the consequence of the default of another. The concept and dimension (dollars) of exposure is the same for all links: it is the total loss that one institution would suffer if a given counterparty defaulted. 

We develop a modification of DebtRank for bipartite networks starting with extending the notion of direct exposures to \emph{indirect exposures} that arise from overlapping portfolios. Let us consider a network of $b$ banks and $m$ assets, and let us denote its equity by $C_i$, the number of shares of asset $a$ owned by bank $i$ by $S_{ia}$, the total number of outstanding shares of asset $a$ by $N_a$, and the price of asset $a$ by $p_a$ respectively. For simplicity we omit the time arguments $t$ for the remainder of the section. 

As a measure of the direct impact of banks on assets we define the matrix
\begin{equation}
	W^{\prime}_{ia}=\frac{S_{ia}}{N_a},
	\label{linear_impact}
\end{equation}
i.e. we assume the impact of bank $i$ on asset $a$ is proportional to the fraction of shares owned by the bank.

Alternatively, we incorporate absorption effects of financial markets into our analysis. When a bank fails, financial markets may have a limited capacity to absorb assets which are sold. Following \citet{Schnabel:2004aa}, \citet{Cifuentes:2005aa} and \citet{Gai:2010aa}, the price of an asset, $p$, is given by
\begin{equation}
	p=e^{-\alpha x}\quad,
\end{equation}
where $x > 0$ is the fraction of the assets sold onto the market (if assets are not being sold onto the market, $p=1$). $\alpha$ may be calibrated so that e.g. the asset price falls by 10\% when one tenth of assets has been sold. To incorporate absorption effects into our analysis, we change \cref{linear_impact} to 
\begin{equation}
	W^{\prime \prime}_{ia}=1-e^{-\alpha W^{\prime}_{ia}}\quad.
	\label{nonlinear_impact}
\end{equation}
Note that with \cref{nonlinear_impact}, $X^{\rm OP}_{ij}$ can no longer be defined as the weighted bank projection of the bipartite network of banks and assets $S_{ia}$. $X^{\rm OP}_{ij}$ defined with \cref{nonlinear_impact} is therefore, in general, not symmetrical.

With the direct impact of banks on assets, we define the indirect exposure of bank $i$ to bank $j$ from overlapping portfolios as
\begin{equation}
	\label{indirect_exposure} X^{\rm OP}_{ij}=\sum_{a} W^{\prime}_{ja} p_a S_{ai} = \sum_{a} \frac{p_{a}}{N_a} S_{ia} S_{aj},
\end{equation}
where $W^{\prime}_{ja}$ is the impact of bank $j$ on assets $a$ and $p_a S_{ai}$ is the current value of assets $a$ in $i$'s investment portfolio. That is, we assume the impact of bank $i$ on bank $j$ is proportional to the impact of bank $j$ on assets $a$ and to the current value of assets $a$ in $i$'s investment portfolio. 

Note that $X^{\rm OP}_{ij}$ is the appropriately weighted bank projection of the weighted bipartite network of banks and assets $S_{ia}$. Assuming the impact of bank $i$ on asset $a$ is proportional to the fraction of shares owned by the bank, $X^{\rm OP}_{ij}$ is symmetrical and the diagonal elements are non-zero, even though the bipartite network itself has, by definition, no self-loops. Diagonal elements represent the self-inflicted loss of a bank from (rapidly) liquidating its portfolio (market impact). This loss will be high if bank $i$ holds a large fraction of asset $a$ in its portfolio, and is negligible if $i$ holds only a small fraction of asset $a$.

To consider contagion from asset liquidation we calculate the DebtRank of the indirect exposure network $X^{\rm OP}_{ij}$, 
\begin{equation}
	R^{\rm OP}_i:=R^{\rm OP}_i(X^{\rm OP}_{ij},C_{i},v^{\rm OP}_{i}) 
\end{equation}
where $C_{i}$ is $i$'s equity and $v^{\rm OP}_{i}$ $i$'s economic value. Given the current value of assets $a$ in $i$'s investment portfolio, we define its economic value as 
\begin{equation}
	\label{ecovalue_op} v^{\rm OP}_{i}=\frac{\sum_{a} p_a S_{ia}}{\sum_{j} \sum_{a} p_a S_{ja}} \quad, 
\end{equation}
i.e. the fraction of $i$'s investment portfolio from the total investment portfolios of all banks. Note that if $\sum_{i}S_{ia}=N_a$, the definition of the economic value in \cref{ecovalue_op} is equal to the definition in \cref{ecovalue}.

$R^{\rm OP}_i$ measures the fraction of the total economic value ($V^{\rm OP}=\sum_{i} \sum_{a} p_a S_{ia}$) that is affected by the distress of a bank $i$ from indirect exposure, i.e. from overlapping portfolios.

\subsection{Quantification of systemic risk in multi-layer networks} \label{multidr}
We consider direct exposures and indirect exposures in the framework of a multi-layer network of exposure networks. In our case we study a multi-layer network consisting of two layers: direct exposures and indirect exposures. Direct exposures represent financial exposures from holding ``deposits \& loans'' (DL), ``derivatives'' (deri), ``securities'' (secu), and ``foreign exchange'' (FX). Indirect exposures result from overlapping portfolios as defined in \cref{indirect_exposure}.

DebtRank values can also be computed for multi-layer networks \citep{Poledna:2015aa}. Specifically, DebtRank values can be computed for each layer of a multi-layer network separately, or for all layers combined (from the combined exposure network $X_{ij}^{\rm comb} = \sum_{\alpha} X_{ij}^{\alpha}$). We refer to the DebtRank of the combined exposure network as $R_i^{\rm comb}$ and the total economic value of the combined exposure network is given by $V^{\rm comb}=\sum_{\alpha} V^{\alpha}=V^{\rm direct}+V^{\rm OP}$. 

To allow a comparison of $R_i^{\alpha}$ between different layers, $R_i^{\alpha}$ must be shown as a percentage of the combined total economic value $V^{\rm comb}$. The normalized DebtRank for layer $\alpha$ is therefore defined in \citep{Poledna:2015aa} as 
\begin{equation}
	\label{debtrank_layer} \hat R_i^{\alpha} = \frac{V^{\alpha}}{V^{\rm comb}} R_i^{\alpha} \quad, 
\end{equation}
where $V^{\alpha}$ is the respective total economic value of layer $\alpha$.

\subsection{Quantification of systemic risk at the country level} \label{sr_profile_sec} 
Using the methodology developed in \citet{Poledna:2015aa}, we estimate systemic risk at the country level. The {\em SR-profile} of a country is defined as the rank-ordered normalized DebtRank $\hat R_i^{\alpha}$ for all financial institutions in a country. The SR-profile shows the distribution of systemic importance across institutions throughout a country. The institution with the highest systemic importance is shown to the very left. The {\rm average DebtRank} is used to capture the SR of the entire economy (with $b$ institutions), 
\begin{equation}
	\bar R ^{\alpha} = \frac{1}{b} \sum_{i=1}^{b} \hat R_i^{\alpha} \quad . 
\end{equation}
For the combined network, $\hat R_i^{\alpha}$ is replaced by $R_i^{\rm comb}$, and we write $\bar R ^{\rm comb}$ for the combined average DebtRank. Note that $\bar R ^{\alpha}$ depends on the network topology of the various layers (or the combined network) only and is independent of default probabilities, recovery rates, or other variables. 

The precise meaning of the DebtRank as the fraction of the total economic value in a network also allows us to define of the {\em expected systemic loss} for the entire economy, which is the size of the loss, multiplied by the probability of that loss occurring \citep{Poledna:2014aa}. Taking into account {\em all} possible combinations of defaulting and surviving institutions and by assuming independent equal loss probabilities of individual institutions, the expected loss for an economy of $b$ institutions can be written as a combinatorial expression 
\begin{equation}
	{\rm EL}^{\rm syst} = V \sum_{S \in \mathcal{P}(B)} \prod_{i \in S} p_i \prod_{j \in B \setminus S} (1-p_j) \: R_S \quad, \label{totEL_normalized} 
\end{equation}
where $p_i$ is the probability of default of institution $i$, and $(1-p_j)$ the survival probability of $j$, $\mathcal{P}(B)$ is the powerset of the set of financial institutions $B$, and $R_S$ is the DebtRank of the set $S$ of nodes initially in distress. The DebtRank of the empty set of nodes initially in distress $R_{\emptyset}$ is defined to be zero. 

It is immediately clear that \cref{totEL_normalized} is computationally feasible only for situations with relative small number of financial institutions. Computing the powerset and calculating DebtRanks for all possible combinations of large financial networks is not feasible. In \citet{Poledna:2015aa} a practical approximation for \cref{totEL_normalized} was derived, 
\begin{equation}
	{\rm EL}^{\rm syst} \approx V \sum_{i=1}^{b} p_i \, R_i \quad, \label{totEL} 
\end{equation}
which is perfectly valid if the default probabilities are small ($p_i \ll 1$), and/or the interconnectedness is low ($R_i \approx v_i$). For details see \citet{Poledna:2015aa}.

\subsection{Quantification of systemic risk of individual exposures}
Following \citet{Poledna:2014aa} and \citet{Poledna:2015aa}, we estimate the marginal SR of individual daily exposures. In particular we compare single exposures of a given size to their marginal SR. The marginal SR of an individual exposure, $\Delta X_{kl}$ (matrix with precisely one nonzero element for the exposure between $k$ and $l$) on ${\rm EL}^{\rm syst}$ is the difference of total expected systemic loss, 
\begin{multline}
	\Delta {\rm EL}^{\rm syst}\bigg|_{\Delta X_{kl}} =   \sum_{i=1}^{b} p_i \left[ V(X_{ij}+\Delta X_{kl})  \,
	R_i(X_{ij}+\Delta X_{kl},C_i) \,   \right.  \\ \left. - V(X_{ij}) \, R_i(X_{ij},C_i) \right] \, \quad, \label{marginal_effect_simple} 
\end{multline}
where $R_i(X_{ij}+\Delta X_{kl},C_i)$ is the DebtRank and $V(X_{ij}+\Delta X_{kl})$ the total economic value of the exposure network without the specific exposure $\Delta X_{kl}$. Clearly, a positive $\Delta {\rm EL}^{\rm syst}$ means that the change in exposures $\Delta X_{kl}$ increases total SR. In general, this risk is borne by the public.

\section{Data} \label{data} 
The data used for this work is derived from a database on exposures at the Mexican Central Bank, built and operated with the specific purpose of studying contagion and SR. This database is maintained by the statistics unit at the financial stability general directorate at this institution. The statistics unit under the financial stability general directorate at Banco de Mexico gathers information and cross-validates it by using  daily, weekly and monthly regulatory reports, which are used for regulatory and supervisory purposes. An illustrative and important example is the case of the daily regulatory reports known as `operaciones de captaci\'on e interbancarias en moneda nacional y udis' (OCIMN), and `operaciones de captaci\'on e interbancarias en moneda extranjera' (OCIME). These reports contain every single funding transaction on a daily basis in local and foreign currency, which are used to compute the daily funding costs for each bank. From these two  regulatory reports it is possible to compute the exact daily unsecured exposures between banks, as well as more broadly, those between financial institutions like investment banks, brokerage houses, mutual funds and pension funds. In \citet{Solorzano-Margain:2013aa}, a stress testing study was carried out using an (extended set) of these exposures. Given the confidential nature of these transactions, data is kept under strict access control and can only be used for regulatory, supervisory and financial stability purposes.

The present work is based on transaction data that is converted to bilateral exposures. Direct and indirect exposures are obtained in the following way. 

\subsection{Direct exposures}

This paper uses data owned by the central bank which comprises exposures arising from different markets and transactions. From this data it is possible to compute gross exposures. The risks considered for building the exposures database include: issuer risk, counterparty risk, credit risk and settlement risk. To compute direct exposures related to issuer risk, information on securities holdings is used. The same information is also used to compute indirect exposures as will be explained later. To compute counterparty risk information on net positions from repos and derivatives is used. The credit risk component is computed by using data on unsecured deposits and loans. Finally, the settlement risk component is computed by using data on foreign exchange transactions.
In most of the previous works, the only type of exposures which were considered was on the unsecured interbank market; nevertheless, this has changed and more jurisdictions are extending the information on the types of exposures in order to perform contagion risk studies. In the Mexican case, the most important type of exposures in terms of amount and SR are exposures that are related to issuer risk, i.e., exposures related to the cross holdings of securities among banks. 

\subsection{Indirect exposures}

The data used for the indirect exposures comes from a database that contains the daily position on securities of each bank and brokerage house. For each day it is possible to determine exactly which securities are held by each bank at the level of individual security and the exact number of them. The number of outstanding securities can be computed; the market price is also obtained from the database. 

Banks hold several types of securities: securities issued by non-bank companies, securities issued by commercial and development banks. Given the unique security  ID, which is part of the database, we can obtain information on the currency in which the security was issued, the issuer and the number of securities issued. Further information on the specific characteristics of each security, like maturity, coupon type, coupon payments, etc. is available. With this information we represent the financial system as the bipartite network $S_{ia}(t)$ of banks and assets at a particular day $t$. The dimensions of the bipartite network vary depending on the number of of outstanding securities held by banks on a particular day.

The vector of prices $p_a(t)$ for the securities is obtained form a database at the Mexican central bank which includes also other relevant aspects for each security. This database is used to compute the number securities issued for each asset ID. 

Finally, balance sheet data on 43 Mexican banks under study is also available, in particular the capitalization, measured on a monthly scale. 

\section{Results} \label{results} 
\subsection{Overlapping portfolios in the Mexican banking system}
\begin{figure*}
	\centering 
	\includegraphics[width=.99\textwidth]{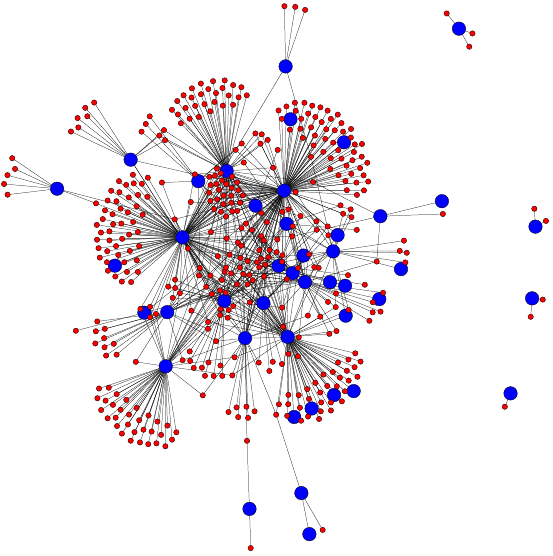} 
	\caption{Bank-asset bipartite network of the Mexican financial system. Nodes in the network represent banks (blue) and assets (red). Links between an asset and a bank exist if the bank holds the asset in its portfolio.} \label{bipartite} 
\end{figure*}

In \cref{bipartite} we visualize overlapping portfolios in the Mexican financial system as a bipartite network. Nodes in the bipartite network represent banks (blue) and assets (red). Links between an asset and a bank exist if a bank holds the asset in its portfolio. 

\begin{figure}
	\centering 
	\includegraphics[width=.25\textwidth]{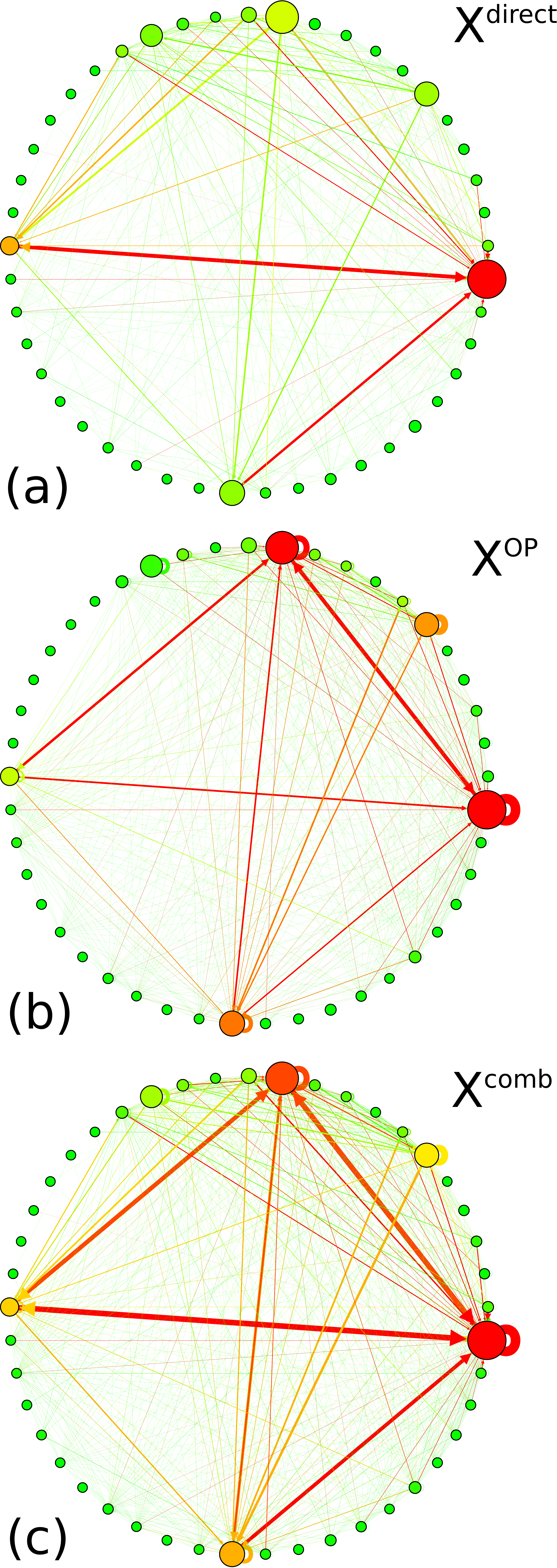} 
	\caption{Multi-layer banking network of Mexico. (a) network of direct exposures from derivatives $X_{ij}^{\rm deri}(t)$, foreign exchange exposures $X_{ij}^{\rm FX}(t)$ and deposits \& loans $X_{ij}^{\rm DL}(t)$, (b) network of indirect exposures from overlapping portfolios $X_{ij}^{\rm OP}(t)$, (c) combined banking network $X_{ij}^{\rm comb}(t)$. The network of indirect exposures is the bank-projection of the bipartite network $S_{ia}(t)$ of banks and assets. Nodes (banks) are colored according to their systemic importance $R_i^{\alpha}$ in the respective layer (see \cref{multidr}): from systemically important banks (red) to systemically safe (green). Node size represents banks' total assets. Link width is the exposure size between banks (re-scaled), link color is taken from the counterparty.} \label{bnetwork} 
\end{figure}

\Cref{bnetwork} shows different layers of exposure of the Mexican financial system. The network of direct exposures from derivatives, foreign exchange exposures, and deposits \& loans is seen in the top layer (\cref{bnetwork}(a)). Nodes are shown at the same position in all layers. Node size represents the size of banks in terms of total assets. Nodes $i$ are colored according to their systemic importance, as measured by the DebtRank, $R_i^{\alpha}(t)$, in the respective layer (see \cref{multidr}). Systemically important banks are red; unimportant ones are green. The width of links represents the size of the exposures in the layer; link color is the same as the counterparty's node color (DebtRank). 

\Cref{bnetwork}(b) shows the network of indirect exposures from overlapping portfolios $X_{ij}^{\rm OP}(t)$ visualized as the bank-projection of the bipartite network $S_{ia}(t)$ of banks and assets shown in \cref{bipartite}. Diagonal elements represent the loss for a bank itself from liquidating its portfolio and are typically larger than the indirect exposure to other banks with similar portfolios, as can be seen in \cref{bnetwork}(b). 

The total exposure from overlapping portfolios is about three times larger ($\sum_{i,j} X_{ij}^{\rm OP}(t) \approx 1\e{12}$ Mex\$) than the direct exposures ($\sum_{i,j} X_{ij}^{\rm direct}(t) \approx 3.3\e{11}$ Mex\$). In \cref{bnetwork}(c) we show the combined exposures $X_{ij}^{\rm comb}(t) = X_{ij}^{\rm direct}(t) + X_{ij}^{\rm OP}(t)$.

\begin{figure}
	\centering 
	\includegraphics[width=.49\textwidth]{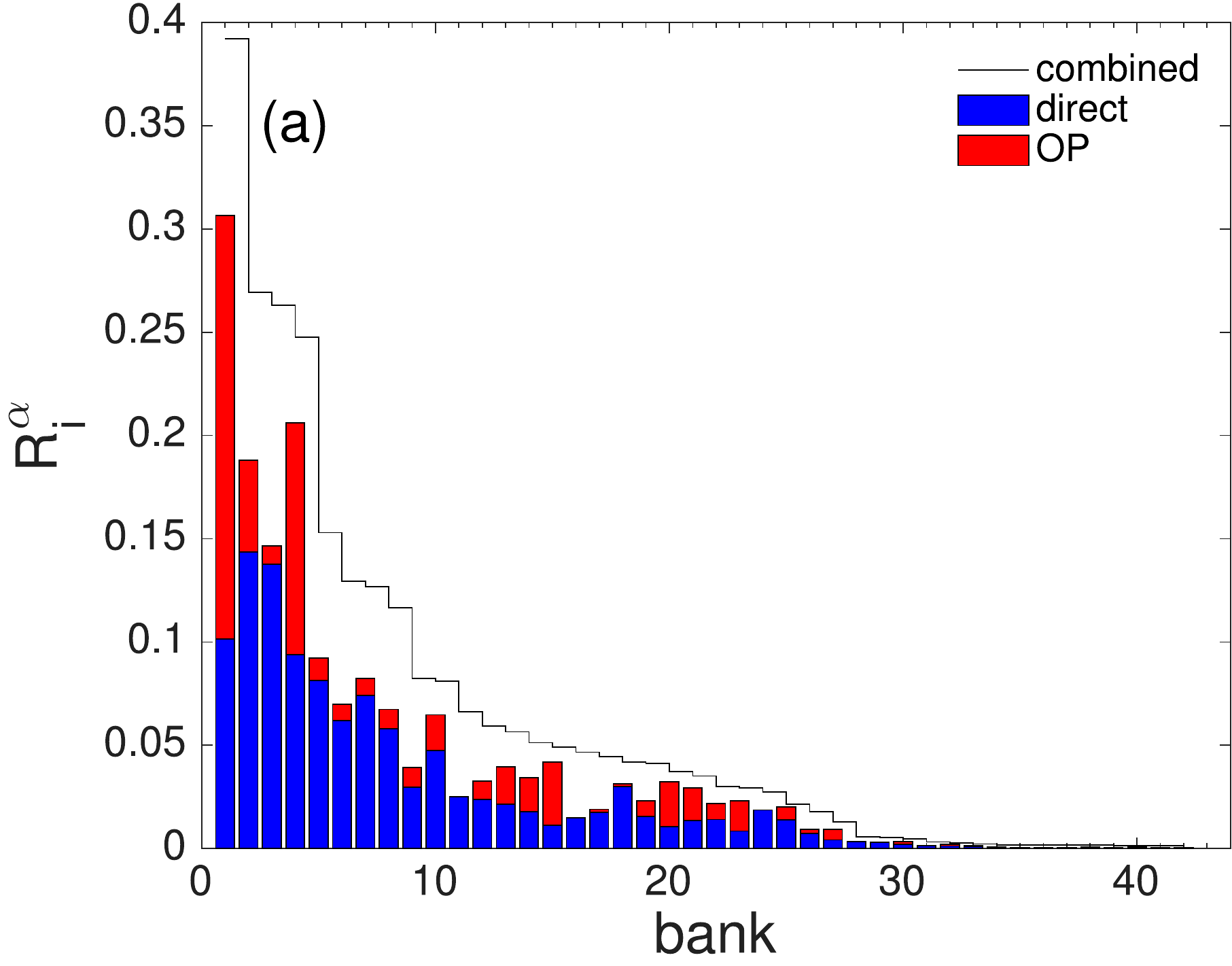} 
	\includegraphics[width=.49\textwidth]{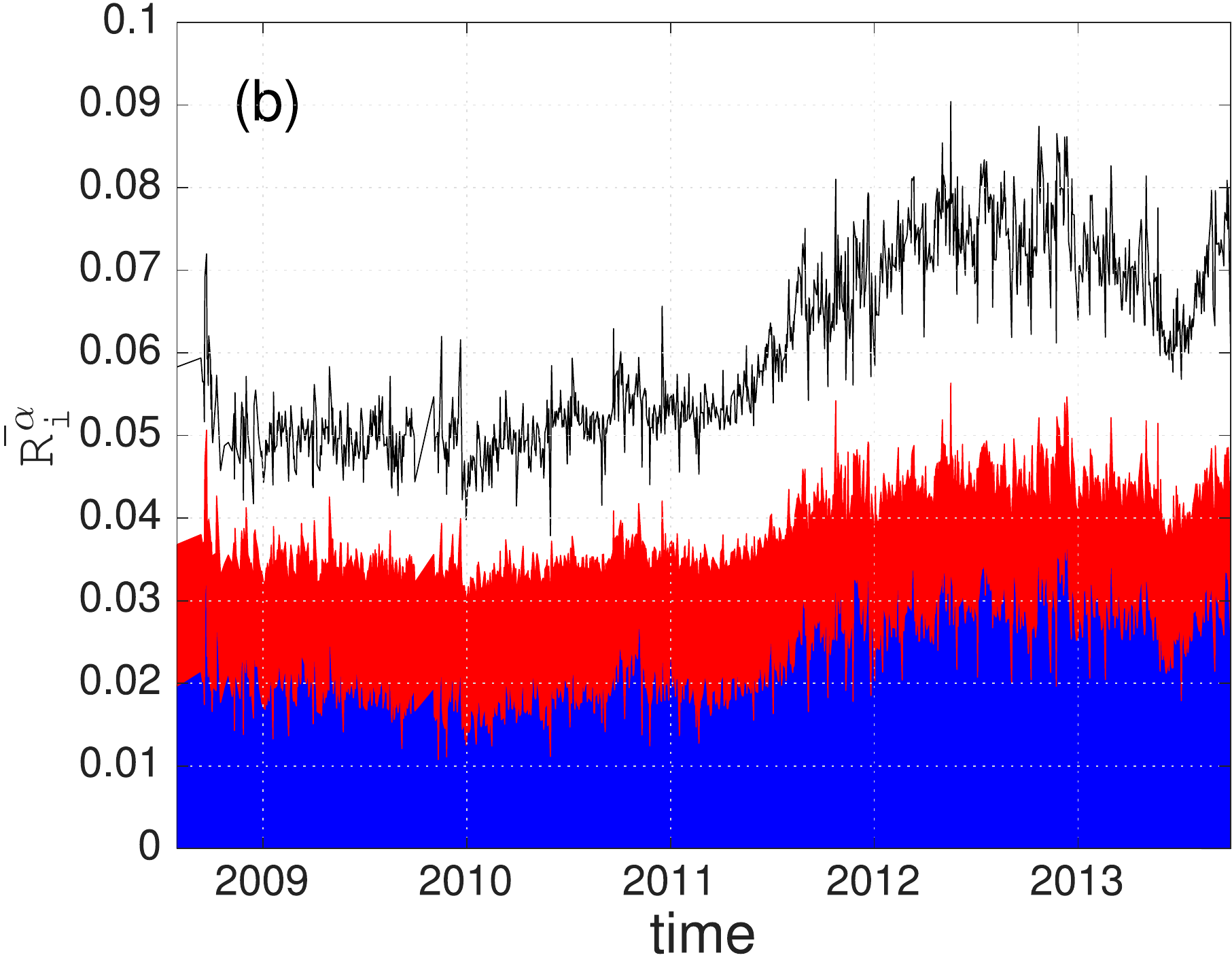} 
	\caption{(a) SR profile for the different layers. Normalized DebtRank $\hat R_i^{\alpha}$ (see \cref{multidr}) from different layers are stacked for each bank. Banks are ordered according to their DebtRank in the combined network from all layers (line). (b) Time series for the average DebtRank $\bar R^{\alpha}(t) = \frac{1}{b} \sum_{i=1}^{b} \hat R_i^{\alpha}(t)$ for all layers from 31 July 2008 to 30 September 2013. The black line shows the average DebtRank for all layers combined $\bar R^{\rm comb}(t)=\frac{1}{b}\sum_{i=1}^{b}R_i^{\rm comb}(t)$.} \label{sr_profile}  
\end{figure}

\Cref{sr_profile}(a) shows the SR-profile for the combined exposures $R_i^{\rm comb}$ (line) and stacked for different layers $\hat R_i^{\alpha}$ (colored bars). Clearly, the SR contribution from direct exposures is smaller than the contributions from indirect exposures. The DebtRank of the combined layers (line) is always larger than the sum of the layers separately, $R_i^{\rm comb}>\sum_{\alpha} \hat R_i^{\alpha}$. 

\Cref{sr_profile}(b) shows the daily {\em average DebtRank} $\bar R$ from 31 July 2008 to 30 September 2013 for the different layers (stacked) and from the combined network (line). As in \cref{sr_profile}(a) the DebtRank of the combined layers is always larger than the combination of the layers separately and the SR contribution from direct exposures is smaller than the contributions from indirect exposures. The contributions of the exposure types are more or less constant over time. 

\begin{figure}
	\centering 
	\includegraphics[width=.49\textwidth]{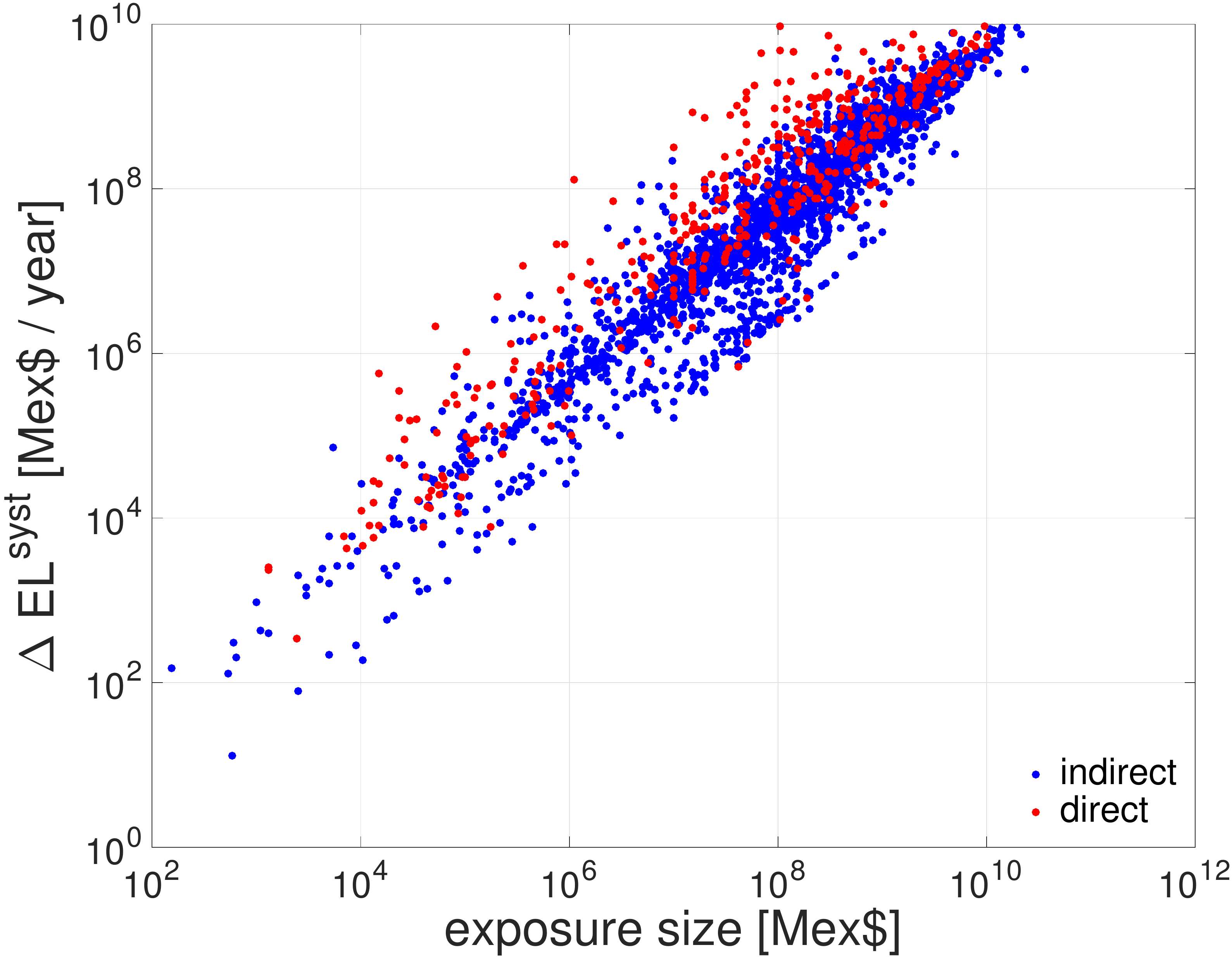} 
	\caption{Marginal increase of expected systemic loss, $\Delta {\rm EL}^{\rm syst}$, versus exposure size, for individual direct and indirect exposures between institutions. Every data point represents an individual exposure $\Delta X_{kl}^{\alpha}$ on a given layer. Every direct and indirect individual exposure between banks is represented by a data point. The marginal SR of individual exposures depends not only on the two  parties involved, but also on the conditions of all nodes in the network.} \label{marginal_effects_figure} 
\end{figure}

In \cref{marginal_effects_figure} we compare the marginal SR of direct and indirect individual exposures with the exposure size. Every direct and indirect individual exposure between banks is represented by a data point. The different layers are distinguished by colors. The marginal SR of individual exposures depend not only on the two parties involved, but also on the conditions of all nodes in the network. Note that the marginal SR of small and medium-size exposures varies by three orders of magnitude. 

\section{Discussion} \label{discussion} 
To a large extent SR arises from indirect interconnections between financial institutions mediated by financial markets. These indirect interconnections occur when financial institutions invest in common assets and are referred to as overlapping portfolios. This work provides, to our knowledge, the first empirical study of systemic risk arising from overlapping portfolios with uniquely identified securities. By having complete information of security holdings of major Mexican financial intermediaries and the ability to uniquely identify securities in their portfolios we represent the Mexican financial system as a bipartite network of securities and financial institutions. In this study we consider overlapping portfolios from securities issued by non-bank companies and securities issued by commercial and development banks (government bonds are excluded).

By generalising DebtRank, a methodology recently introduced to identify systemically important banks in a network of mutual exposures \citep{Battiston:2012aa}, for bipartite networks we extend the notion of systemic importance in financial networks to bipartite networks of financial institutions and securities. This makes it possible to assess SR from overlapping portfolios. Similar to fire sales models we assume a linear price impact, which captures the spirit of most theoretical models of fire sales \citep{Greenwood:2014aa}. Alternatively, we incorporate absorption effects of financial markets into our analysis.

Finally, by representing the financial system as a multi-layer network of direct exposures (default contagion) and indirect exposures (overlapping portfolios), we estimate the mutual influence of different channels of contagion. We show that focusing only on direct exposures underestimates total systemic risk levels by up to 50\%.

\section*{Acknowledgments} The views expressed here are those of the authors and do not represent the views of Banco de M\'{e}xico or the Financial Stability Directorate.

\section*{References} 
\bibliographystyle{elsarticle-harv} 
\bibliography{econophysics}

\clearpage
\appendix 

\section{DebtRank} \label{debtrank_section} DebtRank is a recursive method suggested in \citet{Battiston:2012aa} to determine the systemic importance of nodes in financial networks. It is a number measuring the fraction of the total economic value in the network that is potentially affected by the distress of a node or a set of nodes. For simplicity's sake let us think of the nodes in financial networks as banks. $X_{ij}$ denotes the exposure network at any given moment (loans of bank $j$ to bank $i$), and $C_{i}$ is the capital of bank $i$. If bank $i$ defaults and cannot repay its loans, bank $j$ loses the loans $X_{ij}$. If $j$ does not have enough capital available to cover the loss, $j$ also defaults. The impact of bank $i$ on bank $j$ (in case of a default of $i$) is therefore defined as 
\begin{equation}
	\label{impact} W_{ij} = \min \left[1,\frac{X_{ij}}{C_{j}} \right] \quad. 
\end{equation}
The value of the impact of bank $i$ on its neighbors is $I_{i} = \sum_{j} W_{ij} v_{j}$. The impact is measured by the {\em economic value} $v_{i}$ of bank $i$. Given the total outstanding interbank exposures of bank $i$, $X_{i}=\sum_{j}X_{ji}$, its economic value is defined as 
\begin{equation}
	\label{ecovalue} v_{i}=X_{i}/\sum_{j}X_{j} \quad. 
\end{equation}
To take into account the impact of nodes at distance two and higher, this has to be computed recursively. If the network $W_{ij}$ contains cycles, the impact can exceed one. To avoid this problem an alternative was suggested in \citet{Battiston:2012aa}, where two state variables, $h_{\rm i}(t)$ and $s_{\rm i}(t)$, are assigned to each node. $h_{\rm i}$ is a continuous variable between zero and one; $s_{\rm i}$ is a discrete state variable for three possible states, undistressed ($U$), distressed ($D$), and inactive ($I$), $s_{\rm i} \in \{U, D, I\}$. For a financial networks with $B$ nodes, the initial conditions with $S \subseteq B$ nodes initially in distress are $h_{i}(1) = \Psi \, , \forall i \in S ;\; h_{i}(1)=0 \, , \forall i \not \in S$, and $s_{i}(1) = D \, , \forall i \in S ;\; s_{i}(1) = U \, , \forall i \not \in S$ (parameter $\Psi$ quantifies the initial level of distress: $\Psi \in [0, 1]$, with $\Psi = 1$ meaning default). The dynamics of $h_i$ is then specified by 
\begin{equation}
	h_{i}(t) = \min\left[1,h_{i}(t-1)+\sum_{j\mid s_{j}(t-1) = D} W_{ ji}h_{j}(t-1) \right] \quad. 
\end{equation}
The sum extends over these $j$, for which $s_{j}(t-1) = D$, 
\begin{equation}
	s_{i}(t) = 
	\begin{cases}
		D & \text{if } h_{i}(t) > 0; s_{i}(t-1) \neq I ,\\
		I & \text{if } s_{i}(t-1) = D , \\
		s_{i}(t-1) & \text{otherwise} \quad. 
	\end{cases}
\end{equation}
The DebtRank of the set $S$ (set of nodes in distress at time $1$), is $R^{\prime}_S = \sum_{j} h_{j}(T)v_{j} - \sum_{j} h_{j}(1)v_{j}$, and measures the distress in the system, excluding the initial distress. If $S$ is a single node, the DebtRank measures its systemic importance on the network. The DebtRank of $S$ containing only the single node $i$ is 
\begin{equation}
	\label{debtrank} R^{\prime}_{i} = \sum_{j} h_{j}(T)v_{j} - h_{i}(1)v_{i} \quad. 
\end{equation}
The DebtRank, as defined in \cref{debtrank}, excludes the loss generated directly by the default of the node itself and measures only the impact on the rest of the system through default contagion. For some purposes, however, it is useful to include the direct loss of a default of $i$ as well. The total loss caused by the set of nodes $S$ in distress at time $1$, including the initial distress is 
\begin{equation}
	\label{debtrank_self} R_S = \sum_{j} h_{j}(T)v_{j} \quad. 
\end{equation}

\end{document}